\documentclass{iopart}
\usepackage{graphicx}
\begin{document}
\title{Direct photons at LHC}

\author{A. H. Rezaeian$^1$, B. Z. Kopeliovich$^1$$^-$$^3$, H. J. Pirner$^3$ and Iv\'an Schmidt$^1$}

\address{$^1$ Departamento de F\'\i sica y Centro de Estudios Subat\'omicos, Universidad
T\'ecnica Federico Santa Mar\'\i a, Casilla 110-V, Valpara\'\i so,
Chile   \\
$^2$Joint Institute for Nuclear Research, Dubna, Russia \\
$^3$ Institute for Theoretical Physics, University of Heidelberg,
Philosophenweg 19, D-69120 Heidelberg, Germany }

\ead{Amir.Rezaeian@usm.cl}

\begin{abstract}
 The DGLAP improved color dipole approach provides a good description of
data for inclusive direct photon spectra at the energies of RHIC and
Tevatron.  Within the same framework we predict the transverse momentum
distribution of direct photons at the CERN LHC energies. 
 \end{abstract}

\section{Introduction} Direct photons, i.e. photons not from hadronic
decay, provide a powerful probe for the initial state of matter created in
heavy ion collisions, since they interact with the medium only
electromagnetically and therefore provide a baseline for the
interpretation of jet-quenching models. The primary motivation for
studying the direct photons has been to extract information about the
gluon density inside proton in conjunction with DIS data. However, this
task has yet to be fulfilled due to difference between the measurement and
perturbative QCD calculation which is difficult to explain by
altering the gluon density function (see Ref.~\cite{re1} and references therein). We have recently shown that the color
dipole formalism coupled to DGLAP evolution is an viable alternative to
the parton model and provided a good description of inclusive photon and
dilepton pair production in hadron-hadron collisions \cite{re1}. Here we
predict the transverse momentum spectra of direct photons at the LHC
energies $\sqrt{s}=5.5$ TeV and $14$ TeV within the same framework.

\section{Color dipole approach and predictions for LHC}
\begin{figure}[!t]
       \centerline{\includegraphics[width=11 cm] {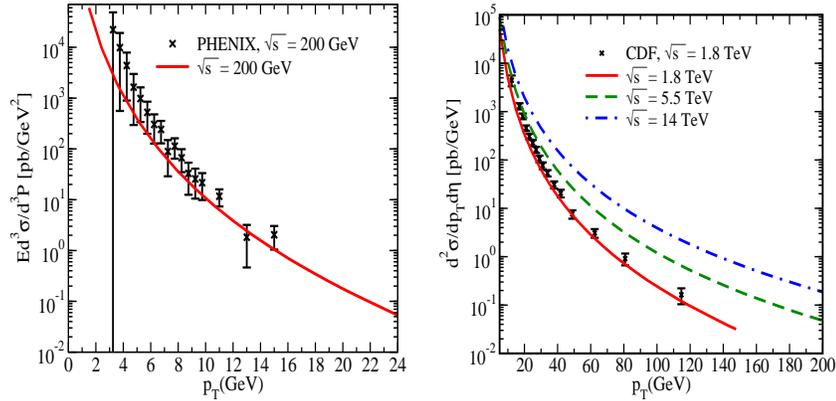}}
       \caption{ Direct photon spectra obtained from GBW-DGLAP dipole
       model at midrapidity for RHIC, CDF and LHC energies. Experimental data (right)
       are for inclusive isolated photon from CDF experiment for $|\eta|<0.9$ at $\sqrt{s}=1.8$ TeV \cite{p4} and (left)
       for direct photon at $\eta=0$ for RHIC energy $\sqrt{s}=200$ GeV \cite{p3}.  The error bars are
       the linear sum of the statistical and systematic
       uncertainties. \label{fig-1}}
\end{figure}

 Although in the process of electromagnetic bremsstrahlung by a quark no
dipole participates, the cross section can be expressed via the more
elementary cross section $\sigma_{q\bar{q}}$ of interaction of a $\bar qq$
dipole. For the dipole cross section, we employ the saturation model of
Golec-Biernat and W\"usthoff coupled to DGLAP evolution (GBW-DGLAP)
\cite{gbw-d} which is better suited at large transverse momenta. Without
inclusion of DGLAP evolution, the direct photon cross section is
overestimated \cite{re1}. In Fig.~\ref{fig-1}, we show the GBW-DGLAP
dipole model predictions for inclusive direct photon production at
midrapidities for RHIC, CDF and LHC energies. We stress that the
theoretical curves in Fig.~\ref{fig-1}, are the results of a parameter
free calculation. Notice also that in contrast to the parton model,
neither $K$-factor (NLO corrections), nor higher twist corrections are
to be added. No quark-to-photon fragmentation function is needed
either.  Indeed, the phenomenological dipole cross section is fitted
to DIS data and incorporates all perturbative and non-perturbative
radiation contributions. For the same reason, in contrast to the
parton model, in the dipole approach there is no ambiguity in defining
the primordial transverse momentum of partons. Such a small purely
non-perturbative primordial momentum does not play a significant role
for direct photon production at the given range of $p_{T}$ in
Fig.~\ref{fig-1}. Notice that the color dipole picture accounts only
for Pomeron exchange from the target, while ignoring its valence
content. Therefore, Reggeons are not taken into account, and as a
consequence, the dipole is well suited mainly for high-energy
processes. As our result for RHIC and CDF energies indicate, we expect
that dipole prescription to be at work for the LHC energies. At the
Tevatron, in order to reject the overwhelming background of secondary
photons isolation cuts are imposed \cite{p4}. Isolation conditions are
not imposed in our calculation. However, the cross section does not vary by more
than $10\%$ under CDF isolation conditions \cite{re1}.  One should also notice that the
parametrizations of the dipole cross section and proton structure
function employed in our computation have been fitted to data at
considerably lower $p_{T}$ values \cite{re1}.

\ack
This work was supported by Fondecyt grants 1070517 and 1050519 and by DFG grant PI182/3-1.

\end{document}